# Fundamental physics activities with pulsed neutron at J-PARC(BL05)

Kenji MISHIMA[1], Shogo AWANO[2], Yasuhiro FUWA[3], Fumiya Goto[2], Christopher C. Haddock[4], Masahiro Hino[5], Masanori Hirose[6], Katsuya Hirota[2], Sei Ieki[7], Sohei Imajo[2], Takashi Ino[1], Yoshihisa Iwashita[3], Ryo Katayama[3], Hiroaki Kawahara[2], Masaaki Kitaguchi[8], Ryunosuke Kitahara[3], Jun Koga[9], Aya Morishita[9], Tomofumi Nagae[6], Naoki Nagakura[7], Naotaka Naganawa[2], Noriko Oi[2], Hideyuki Oide[10], Hidetoshi Otono[11], Yoshichika Seki[12], Daiichiro Sekiba[13], Tatsushi Shima[14], Hirohiko M. Shimizu[2], William M. Snow[4], Naoyuki Sumi[9], Hirochika Sumino[15], Satomi Tada[2], Kaoru Taketani[1], Seiji Tasaki[16], Tatsuhiko Tomita[9], Atsuhiro Umemoto[2], Takahito Yamada[7], Satoru Yamashita[16], Mami Yokohashi[2], Tamaki Yoshioka[11]

[1] High Energy Accelerator Research Organization
[2] Department of Physics, Nagoya University
[3] Institute for Chemical Research, Kyoto University
[4] Department of Physics, Indiana University
[5] KURRI, Kyoto University
[6] Department of Physics, Kyoto University
[7] Department of Physics, University of Tokyo
[8] KMI, Nagoya University
[9] Department of Physics, Kyushu University
[10] INFN-Genova
[11] Research Center for Advanced Particle Physics, Kyushu University
[12] Japan Atomic Energy Agency
[13] Institute of Applied Physics, University of Tsukuba
[14] RCNP, Osaka University
[15] Dept. of Basic Sci., University of Tokyo
[16] Department of Nuclear Engineering, Kyoto University
[17] International Center for Elementary Particle Physics, University of Tokyo
*E-mail: kenji.mishima@kek.jp

"Neutron Optics and Physics (NOP/ BL05)" at MLF in J-PARC is a beamline for studies of fundamental physics. The beamline is divided into three branches so that different experiments can be performed in parallel. These beam branches are being used to develop a variety of new projects. We are developing an experimental project to measure the neutron lifetime with total uncertainty of 1 s (0.1%). The neutron lifetime is an important parameter in elementary particle and astrophysics. Thus far, the neutron lifetime has been measured by several groups; however, different values are obtained from different measurement methods. This experiment is using a method with different sources of systematic uncertainty than measurements conducted to date. We are also developing a source of pulsed ultra-cold neutrons (UCNs) produced from a Doppler shifter are available at the unpolarized beam branch. We are developing a time focusing device for UCNs, a so called "rebuncher", which can increase UCN density from a pulsed UCN source. At the low divergence beam branch, an experiment to search an unknown intermediate force with nanometer range is performed by measuring the angular dependence of neutron scattering by noble gases. Finally the beamline is also used for the research and development of optical elements and detectors. For example, a position sensitive neutron detector that uses emulsion to achieve sub-micrometer resolution is currently under development. We have succeeded in detecting cold and ultra-cold neutrons using the emulsion detector.

**1. Introduction**

BL05 (NOP) is a beamline to carry out fundamental physics experiments using neutrons. The beamline is divided upstream into three branches to conduct different experiments in parallel [1], [2]. The polarized beam branch uses magnetic supermirrors bender containing magnetized materials to produce highly polarized neutrons for the neutron lifetime and asymmetry experiment. The unpolarized beam branch produces the highest energy and most intense neutron flux in these branches with a supermirror bender and will be used to produce ultra-cold neutrons (UCNs) using a Doppler shifter. The low-divergence beam branch, which produces small divergence but a high luminosity neutron beam, is used for neutron small angle scattering to search for a new medium-range force. In this letter, we will introduce the present results and on-going activities at BL05 (NOP) beamline.

**2. Measurement of neutron lifetime**

A neutron decays into a proton, an electron, and an antineutrino with a lifetime of ~880 s. The neutron lifetime is an important parameter for the unitarity of the CKM matrix and primordial big bang nucleosynthesis. However, recently reported measurements of the neutron lifetime deviate significantly (8.4 s or 4.0 σ) from the systematic errors attributed to measurement methods [3], [4]. To solve the problem, an experiment to measure the neutron lifetime with a pulsed beam is on-going in the polarized branch of the BL05 (NOP) beamline. In this experiment, the neutron lifetime is measured as the ratio of the electron events to the $^3$He(n,p)$^3$H events caused by precisely doped $^3$He gas using a time projection chamber (TPC) [5]. In this method, the neutron lifetime $\tau_n$ is derived as

$$\tau_n = \frac{1}{\rho_{He}\sigma_{He}(v_0)v_0} \frac{(S_{He}/\epsilon_{He})}{(S_\beta/\epsilon_\beta)} \quad (1),$$

where $S_{He}$, $S_\beta$ and $\epsilon_{He}$, $\epsilon_\beta$ are counts and detection efficiencies of $^3$He(n,p)$^3$H and beta decay, respectively. $\rho_{He}$ is number density of $^3$He, $\sigma_{He}(v_0)$ is the cross section of the $^3$He(n,p)$^3$H reaction at a neutron velocity of $v_0$, which is 2,200 m/s.

The TPC has 1 m length with 30 cm height and width, whose inner surface was lined by $^6$LiF/PTFE tiles to capture scattered neutrons without gamma rays. The TPC response was calibrated using the 5.9 keV Kα X-ray from $^{55}$Fe on a rotating stage [6]. Neutron bunches of 40 cm length are shaped by a spin flip chopper (SFC) and delivered to the TPC with a repetition rate of 5 bunches per pulse, which means 125 bunches per second.

*2.1 Analysis*

Events were selected by time of flight (TOF), where the neutron bunches were entirely in the TPC and by subtracting background in a region where no neutron are present. Backgrounds caused by γ-rays from SFC were subtracted by closing a $^6$LiF shutter between SFC and TPC, which are alternately taken with shutter open. After subtracting the background signal, events were selected by pulse height and topologies to $^3$He(n,p)$^3$H or neutron beta decay events. A Monte Carlo simulation was used to check the validity and evaluate the efficiencies [7]. The number density of $^3$He, $\rho_{He}$, doped as partial pressure of ~100 mPa in TPC working gas (He 85 kPa + $CO_2$ 15 kPa) was determined by introducing $^3$He in an initial volume, where the volume ratio to the TPC vacuum chamber was measured with a static expansion method using high precision transducers. The present systematic uncertainties of a gas fill measured in 10 days with 200 kW of beam power are listed in Table 1.

Table 1. Uncertainty budget of a gas fill measured 2016 (preliminary).

| | Source | Uncertainty [%] | | Source | Uncertainty [%] |
|---|---|---|---|---|---|
| | γ ray spectra from LiF | 0.04 | | Number density of $^3$He | 0.39 |
| $S_\beta$ | $CO_2$ contamination | 0.27 | $\rho_{He}$ | Deformation of TPC by pressure | 0.31 |
| | $^3$He contamination | 0.2 | | Deformation of TPC by temperature | 0.04 |
| | Pile up | 0.36 | | Temperature non-uniformity in TPC | 0.22 |
| | SFC contrast effect | 0.02 | | $S_\beta$ (statistic) | 2.59 |
| $\epsilon_\beta$ | Cut efficiency for β | 0.92 | | $S_\beta$ (systematic) | 0.49 |
| | Pile up | 0.01 | Subtotal | $S_{He}$ (statistic) | 0.18 |
| $S_{He}$ | $^{14}$N contamination | 0.06 | | $S_{He}$ (systematic) | 0.07 |
| | $^{17}$O contamination | 0.03 | | $\rho_{He}$ | 0.55 |
| | β contamination | 0.03 | | $\sigma_{He}$ | 0.13 |
| $\epsilon_{He}$ | Cut efficiency for $^3$He | 0.008 | $\tau_n$ | total | 2.85 |

## 2.2 Future Upgrades

Upgrades for future experiments are on-going. A large area SFC, which is expected to increase the neutron intensity by a factor 23.5, was prepared[8]. Low power consumption amplifiers with ASIC [9] were prepared for low pressure operation of the TPC to prevent gas induced neutron scattering.

## 2.3 Cross section of $^{14}$N(n,p)$^{14}$C

The experimental apparatus, SFC, TPC, and gas handling system could be used for precise measurement of neutron capture reactions of many nuclei. We have measured the reaction cross section of $^{14}$N(n,p)$^{14}$C, which is known to work as a neutron poison in stellar reaction [10]. This reaction has been previously measured with an accuracy of

1.6%. Our preliminary result was 1.864(8) barn, whose accuracy was 0.4%, consistent with the mean value from previous measurements (Fig.1) [11].

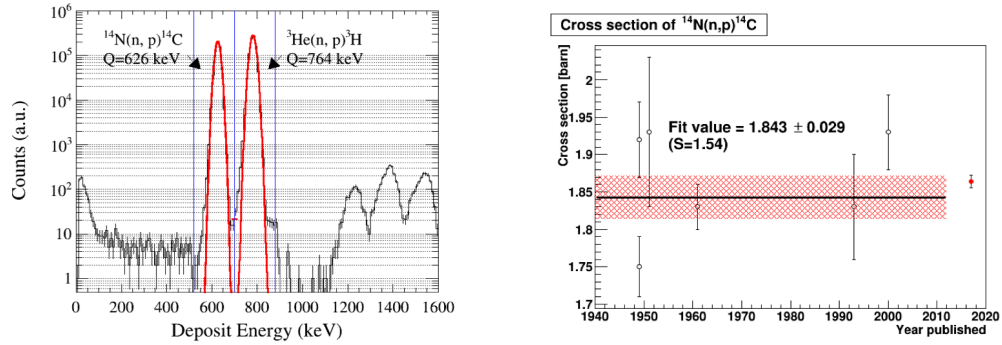

**Fig. 1.** Energy spectrum of $^{14}$N(n,p)$^{14}$C and $^{3}$He(n,p)$^{3}$H reactions(left). Cross sections of $^{14}$N(n,p)$^{14}$C, previous and present (right). The present cross-section (preliminary) was pointed in the last, which was consistent with the mean value of previous results.

## 3. Ultra-cold neutrons

Ultra-cold neutrons (UCNs) are neutrons with kinetic energy that is less than ~250 neV, or velocity of 6.8 m/s [12]. Such low energy neutrons can be stored in a vessel via total reflection on material surfaces. UCNs are also sensitive to magnetic fields (60 neV/T) and gravitational potentials (100 neV/m). Because of the special feature, UCNs are utilized for various precision measurements in fundamental physics.

### 3.1 Doppler shifter for pulsed UCN production

To develop and test UCN devices, a Doppler shifter was installed to provide pulse UCNs at the unpolarized beam branch of the BL05 (NOP) beamline. The Doppler shifter consist of a supermirror of $m = 10$ rotating at 2,000 rpm. Very cold neutrons (VCNs) with velocity of 136 m/s are reflected into the Doppler shifter by monochromatic supermirrors. The rotating mirror with velocity of 68 m/s reflects a VCN every 120 ms, and produces neutrons with velocities that are less than 20 m/s. The reflected neutron intensity at an extraction port was 410 cps with 1 MW equivalent beam power. The UCN intensity and peak UCN density are estimated to be 160 cps and 1.4 /cm$^3$ according to the simulation[13], [14].

### 3.2 UCN rebuncher

For the pulsed UCN source, the UCN density is diluted with time and therefore, the UCN density at the experimental apparatus is significantly smaller than the density at the UCN source. It is possible in principle that controlling the UCN velocities with TOF can focus the pulsed UCNs at a certain point in space and time, thus recovering the UCN density. This method is expected to enhance the UCN density that is used in an experimental apparatus.

We are currently developing a time focusing device for UCN, which we have named "UCN rebuncher"[15]. The UCN rebuncher consists of a gradient dipole magnet and a frequency variable RF spin flipper. As a neutron enters a magnetic field of 1 T, it loses its kinetic energy of 60 neV of kinetic energy. If the spin of the neutron is flipped in the magnetic field adiabatically, the neutron again loses another 60 neV as it leaves the magnetic field. In this way, we can decrease the velocity of neutrons.

The gradient dipole magnet generates a linearly changing magnetic field up to 1.0 T [16]. By changing the RF frequency and synchronizing TOF of the UCN from 6–30 MHz, which correspond to the resonance frequency of 0.2–1.0 T, we can select the UCNs that will flip their spin by choosing the appropriate magnetic field. Well controlled synchronization of the RF frequency using variable condensers makes it possible to focus UCNs at 3.5 m downstream from the rebuncher.

We performed the time focusing experiment in 2017. The experimental setup is shown in Fig. 2. UCN pulses were generated with a frequency of 0.5 Hz by a UCN shutter in front of the source points. TOF spectra of UCN with/without RF modulation is also shown in right. It is observed that UCNs in TOF of 1,000–1200 ms were swept out and focused at 1,250 ms [17].

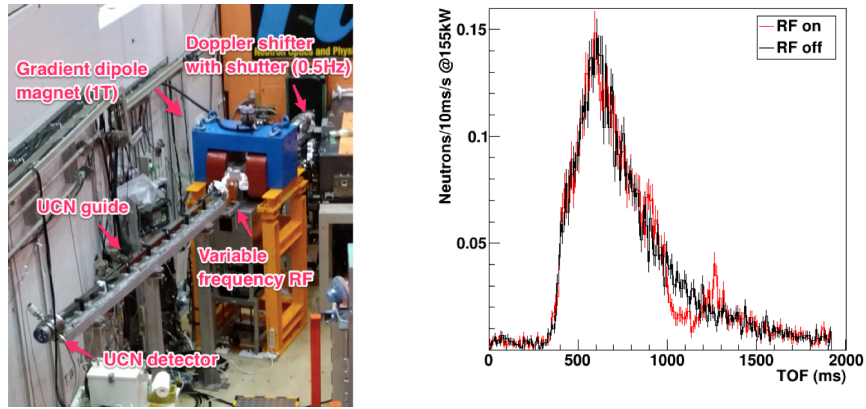

**Fig. 2.** Experimental setup for UCN rebuncher experiment (left). TOF spectra of UCN with/without RF(right). It is observed that UCNs in 1,000–1,200 ms were swept out and focused at 1,250 ms [17].

*3.3 UCN storage*

UCN storage has been used to study reflectivity of UCN on surfaces. UCNs were stored in an electrolytic polished stainless-steel vessel that has two valves to confine UCN for storage and extraction. The maximum measured UCN storage time was 270 s (Fig. 3). An aluminum filter on the exit port at the Doppler shifter shortened the storage time to 180 s. This may occur if the low-energy UCNs below the Fermi potential of Al (54 neV) were filtered, so that the higher energy of the UCNs experienced more collisions at the wall of the vessel. The storage time of 180 sec corresponds to UCN loss rate per bounce ($\eta$) as $\eta \sim 2 \times 10^{-4}$. The increasing the vacuum

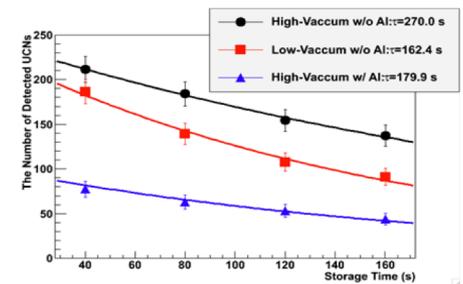

**Fig. 3.** Storage time of UCNs in the stainless storage vessel. The maximum storage time was 270 sec (black). Al filter on the exit of the Doppler shifter shortened the storage time to 180 sec (blue). The worsening of the vacuum (few Pa) also decreased the storage time to 162 sec.

pressure from ~$10^{-1}$ Pa to ~3 Pa also decreased the storage time to 162 s.

**4. Search for an Unknown Intermediate force with Noble gas scattering**

There are four fundamental interactions in Nature: strong, weak, electromagnetic, and gravitational. The first three are unified within the standard model. However, gravity is not described within the standard model framework and its functional form may not be known exactly. Therefore many experiments have been performed to search for possible deviations from the inverse-square law of Newtonian gravity, particularly in the short length region, down to tens of micrometers. Precise neutron scattering measurements from noble gases can test the $1/r^2$ dependence of the gravitational force on even shorter length scales, such as the submicron region. We describe an experiment which probes extra possible forces in the nanometer length scale by measuring neutron scattering from noble gases.

In the simplest case of an as-of-yet unknown particle with a Compton wavelength of $\lambda$, a possible Yukawa force might add to the Newtonian force so that the potential $V(r)$ between masses of $m$ and $M$ can be described as

$$V(r) = -G_N \frac{mM}{r}(1 + \alpha e^{-r/\lambda}) \qquad (2)$$

where $r$ is distance between the masses, and $G_N$ is the gravitational constant. Neutron scattering is a unique probe to measure the short-range forces because it is chargeless and contains low polarizability [18]. Additional forces described as in eq. (2) increase the forward intensity in the angular distribution in neutron scattering as shown in Fig 4. We measure the small angle scattering from Xe gas where extra forces are most sensitive.

The experiment was conducted at the low divergence beam branch at BL05 (NOP). A schematic view of the experimental apparatus containing the gas cell, scattering chamber, and detector is shown in Fig 4. Neutrons pass a gas cell containing a sample gas (Xe, Ar, and He) with pressure of 200 kPa. Neutron scattered by the nuclear and/or new short-range force are measured by position sensitive detectors whose resolution is ~5 mm in the horizontal direction. Nuclear scattering and short-range forces are distinguished by their angular dependences [19], [20]. Up to this point, we have collected ~$2 \times 10^7$ neutron scattering events [21].

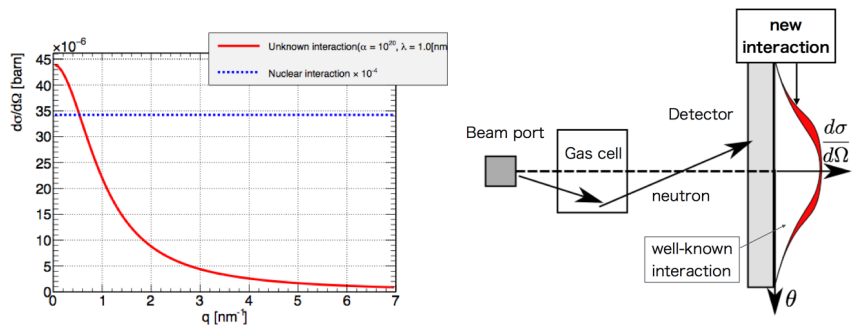

**Fig. 4.** Differential cross section of extra Yukawa force and nuclear interaction(left). Schematic view of the experimental setup of the gas scattering experiment(right).

## 5. Development of a sub-micrometer position sensitive neutron detector using nuclear emulsion

Nuclear emulsion, which is a kind of photographic film, is used as a high sensitivity position tracking detector for particle and nuclear physics. We applied the nuclear emulsion to neutron detection with sub-micrometer spatial resolution. Tracks of ions from a neutron capture reaction, i.e., $^7$Li(n,t)$^4$He and $^{10}$B(n,α)$^7$Li, can be identified by the nuclear emulsion, and it is possible to determine the starting point of ions where the neutron was captured. The resolution of the track identification is inversely proportional to the density of grains in the tracks, which is typically ~1.4 grain/μm for fine-grained nuclear emulsion. Thus, we expect to achieve sub-micrometer spatial resolution [22].

We are developing a $^{10}$B layer type detector as a high spatial resolution neutron detector. Because the position of the neutron reaction points is fixed, we can determine the starting point by extrapolating the grains of a track. This detector is expected to have less spatial resolution than the grain density.

The schematic view of a $^{10}$B layer type detector is shown in Fig. 5. The fabricated detector was exposed to cold neutrons at the low-divergence beam branch and UCNs from the Doppler shifter. After development, we observed tracks in the $^{10}$B detector by neutron capture for both exposures [23]. We are developing an automatic track recognition algorithm for future experiments [24].

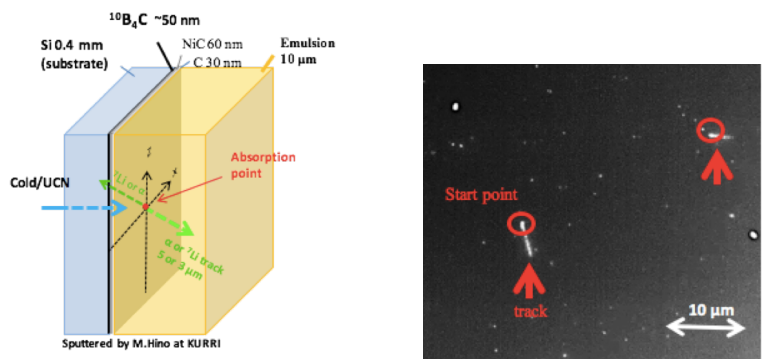

**Fig. 5.** Schematic view of neutron detector of nuclear emulsion(left). 50 nm of $^{10}$B$_4$C layer on Si substrate is covered by nuclear emulsion (10 μm). Tracks observed by UCN exposure(right) [23]. Because the depth of the neutron reaction points is fixed, we can determine the starting point extrapolating by fitting of grains of a track.

## 6. Conclusion

"Neutron Optics and Physics (NOP/ BL05)" beamline is working at MLF in J-PARC for studies of fundamental physics. We have reviewed experiments performed at the beamlines with certain achievements.

The neutron lifetime measurement is on-going at the polarized beam branch and has taken physics data with an expected uncertainty for the 2016 run of ~20 s, and an ultimate goal for the uncertainty of ~1 s. Pulsed ultra-cold neutrons (UCNs) by a Doppler shifter are available at the unpolarized beam branch. We have demonstrated the time focusing using the so-called "rebuncher," which can increase UCN density from a pulsed UCN source. At the low divergence beam branch, we performed an experiment to search for an unknown intermediate force in the nanometer range by measuring the angular dependence of neutron scattering by noble gases. We successfully demonstrated the detection of cold and ultra-cold neutrons using the emulsion detector with a sub-micrometer spatial resolution.

**Acknowledgment**

This work was supported by JSPS KAKENHI JP16H02194, JP26247035, and JP23244047). The neutron experiment at the Materials and Life Science Experimental Facility of the J-PARC was performed under S-type project of KEK (Proposal No. 2014S03) and user programs: neutron lifetime experiment (Proposal No. 2016A0161, 2015A0316, 2014B0271, and 2014A0244), UCN experiments (Proposal No. 2016A0168, 2015A0317, and 2014B0253), gas scattering experiment (Proposal No. 2016B0212, 2016A0078, and 2015A0239), and emulsion experiment (Proposal No. 2016A0213, 2015A0242, and 2014B0270). The ASIC preamplifier was developed with Open-It program.